\documentclass[11pt]{article} 
\usepackage{mystyle-new}
\usepackage{authblk}
\usepackage[normalem]{ulem}
\usepackage[T1]{fontenc}
\usepackage{epsfig,amsmath} 
\usepackage{hyperref}
\usepackage{color}
\usepackage{graphicx}
\usepackage{orcidlink}
\definecolor{red}{rgb}{1,0,0}
\def\lesssim{\ \hbox{\raise 2pt \hbox{$<$} \kern -13pt
                     \lower 3pt \hbox{$\sim$}}\ }
\def\greatersim{\ \hbox{\raise 2pt \hbox{$>$} \kern -13pt
                     \lower 3pt \hbox{$\sim$}}\ }

\def\lsim{\mathrel{\rlap{\lower4pt\hbox{\hskip1pt$\sim$}}
    \raise1pt\hbox{$<$}}}                
\def\gsim{\mathrel{\rlap{\lower4pt\hbox{\hskip1pt$\sim$}}
    \raise1pt\hbox{$>$}}}                

\def\cascade{{\sc Cascade3}}
\def\pythia{{\sc Pythia}}
\def\herwig{{\sc Herwig}}

\def\mcatnlo{{MCatNLO}}

\input epsf.tex
\def\desepsf(#1 width #2){\epsfxsize=#2 \epsfbox{#1}}
\def\kt{\ensuremath{k_{\rm T}}}

\def\pt{\ensuremath{p_{\rm T}}}
\def\PZ{\ensuremath{Z}}

\def\qt{\ensuremath{q_{\rm t}}}
\def\zdyn{\ensuremath{z_{\rm dyn}}}
\def\ptll{\ensuremath{p_{\rm T}(\ell\ell)}}

\newcommand{\alphas}{\ensuremath{\alpha_\mathrm{s}}}

\newcommand{\mdy}{\ensuremath{m_{\small\text{DY}}}}

\newcommand{\PBM}{PB}
\newcommand{\PBset}{{PB-NLO-2018}}

\newcommand{\MCatNLO}{{\sc MadGraph5\_aMC@NLO}}

\newcommand{\GeV}{GeV}
\newcommand{\TeV}{TeV}
\newcommand{\as}{\ensuremath{\alpha_s}}
\newcommand{\softgluon}{NP-gluon}

\newenvironment{tolerant}[1]{\par\tolerance=#1\relax}{ \par }
\usepackage{amsmath,bm}
\usepackage{lineno}

\usepackage{cite,./mcite}
\usepackage{tikz}
\usepackage[symbol]{footmisc}

\newcommand{\dglap}{Gribov:1972ri,Lipatov:1974qm,Altarelli:1977zs,Dokshitzer:1977sg}

\providecommand{\DOI}[1]{\href{http://dx.doi.org/#1}}

\begin{document}

\title{
On the role of soft and non-perturbative gluons  in collinear parton densities and parton shower event generators }

\author[1]{M.~Mendizabal~\orcidlink{0000-0002-6506-5177}}
\affil[1]{Deutsches Elektronen-Synchrotron DESY, Germany}
\author[2]{F.~Guzman\footnote{deceased in Oct 2024}~\orcidlink{0000-0002-7612-1488}}
\affil[2]{InSTEC, Universidad de La Habana, Havanna, Cuba}
\author[1,3,4]{H.~Jung~\orcidlink{0000-0002-2964-9845}}
\affil[3]{II. Institut f\"ur Theoretische Physik, Universit\"at Hamburg,  Hamburg, Germany}
\affil[4]{Elementary Particle Physics, University of Antwerp, Belgium}
\author[1]{S.~Taheri~Monfared~\orcidlink{0000-0003-2988-7859}}

\date{}
\begin{titlepage} 
\maketitle
\vspace*{-10cm}
\begin{flushright}
DESY-23-172\\
\today
\vspace*{7.5cm}
\end{flushright}

\centerline{This paper is dedicated to the memory of Fernando Guzman. }

\abstract{
The Parton Branching method offers a Monte Carlo solution to the DGLAP evolution equations by incorporating Sudakov form factors. In this approach, the Sudakov form factor can be divided into perturbative and non-perturbative components, with the non-perturbative part being analytically calculable under specific conditions.

\begin{tolerant}{5000}
We first examine forward evolution and demonstrate that including soft and non-perturbative gluons  (through the non-perturbative Sudakov form factor) is essential for the proper cancellation of divergent terms in parton density evolution. This non-perturbative component is also important for Transverse Momentum Dependent (TMD) parton distributions, and within the Parton Branching framework, it is constrained by fits to inclusive collinear parton densities.
\end{tolerant}

Additionally, we explore the impact of this non-perturbative Sudakov form factor on backward parton evolution and its effects on parton and hadron spectra originating from initial state showers. Our results show that soft and non-perturbative gluons  significantly influence inclusive distributions, such as Drell-Yan transverse momentum spectra. However, we found that soft and non-perturbative gluons  have a minimal impact on final state hadron spectra and jets.

}

\end{titlepage}

\section{Introduction}
\label{sec:intro}
Calculations based on the DGLAP~\cite{\dglap} evolution of parton densities, combined with hard scattering coefficient functions (or matrix elements) at next-to-leading order (NLO) and next-to-next-to-leading order (NNLO) in the strong coupling, provide a highly successful description of experimental measurements across a broad range of energy scales $\mu$.

The Parton-Branching (\PBM) approach~\cite{Hautmann:2017fcj,Hautmann:2017xtx} offers a solution to the DGLAP equations through an iterative solution of the integral evolution equation. The \PBM\ approach enables detailed analysis of each branching vertex, particularly focusing on the contributions of perturbative and non-perturbative emissions. 
In the context of Transverse Momentum Dependent (TMD) parton densities~\cite{Collins:2017oxh,Jung:2021mox}, and especially within the CSS formalism~\cite{Collins:1984kg}, a non-perturbative Sudakov form factor is introduced. In the \PBM\ approach, this non-perturbative Sudakov form factor is inherently present in inclusive (collinear) parton densities. By fitting collinear parton densities to inclusive experimental data, this non-perturbative Sudakov form factor is determined, and can subsequently be applied to TMD parton densities.

The \PBM -method provides an intuitive bridge between TMD resummation and parton showers in general-purpose Monte Carlo event generators. The \PBM -method explicitly simulates each individual branching, similar to a parton shower. However, unlike conventional parton showers, the \PBM\ TMD parton shower is directly linked to the \PBM\ TMD distributions, with no free parameters. Importantly, once integrated, the \PBM\ TMD leads exactly to the inclusive distribution by construction. The \PBM -approach can be applied to study the effects of soft and non-perturbative gluons, which are often neglected in conventional parton showers.

In this paper, we show explicitly how the Sudakov form factor is obtained from the DGLAP evolution equation and how this form factor can be split into a perturbative and non-perturbative part. We argue that both parts are essential for collinear parton distributions, as neglecting soft and non-perturbative gluons would lead to non-cancellation of singular contribution in cross section calculations at NLO and beyond. 
We also show, that soft and non-perturbative gluons (in the following abbreviated as \softgluon s), while essential for inclusive distributions, do not play a role in final state hadron spectra, if a hadronization model, such as the Lund String model~\cite{Sjostrand:1993yb,Bierlich:2022pfr}, is used.

We claim that including \softgluon s in both parton densities and the parton shower yields a consistent picture with NLO DGLAP calculations, in contrast to the claims in Ref.\cite{Frixione:2023ssx}. Furthermore, we argue that the proper treatment of \softgluon s is essential for obtaining an intrinsic-\kt\ distribution that accurately describes the Fermi motion of partons inside hadrons, independent of the center-of-mass energy of the process. This contrasts with observations\cite{Bubanja:2024puv} from traditional parton shower event generators like \pythia 8 or \herwig, where the  intrinsic-\kt\ distribution is $\sqrt{s}$ dependent.

The paper is organized as follows. In Section~\ref{sec:PBTMD}, we begin by briefly reviewing the basic elements of our calculation  framework, the PB TMD approach. Section~\ref{sec:Sudakov} forms the core of the paper, where we discuss in detail the non-perturbative contributions to the Sudakov form factor and underscore the significance of \softgluon\  emissions in both collinear and transverse momentum dependent parton distributions as well as in cross section calculations. In Section~\ref{sec:Implications}, we discuss the  importance of accurately treating \softgluon\  emissions in initial state showers for partons and their  effect on finals state hadron spectra. Finally, we present our conclusions in Section~\ref{sec:concl}. In the appendices we comment on a few technical aspects.

\section{PB approach as a solution of DGLAP evolution equations}
\label{sec:PBTMD}

The \PBM -method~\cite{Hautmann:2017fcj,Hautmann:2017xtx} provides a solution to the DGLAP~\cite{\dglap} evolution equations. The DGLAP evolution equation for the parton density 
of parton $a$ with momentum fraction $x$ at the scale $\mu$ reads:
\begin{equation}
\label{EvolEq}
 \mu^2 \frac{{\partial }{ f}_a(x,\mu^2)}{{\partial } \mu^2}   =  
 \sum_b
\int_x^{1} {\frac{dz}{z}} \; {P}_{ab} \left(\as(\mu^{2}),z\right)  \;{f}_b\left({\frac{x}{z}},
\mu^{2}\right)   \; ,
\end{equation}
with the regularized DGLAP splitting functions $P_{ab}$ describing the splitting of parton $b$ into a parton $a$. The splitting functions $P_{ab}$ can be decomposed as (in the notation of Ref.~\cite{Hautmann:2017fcj}):
\begin{equation}
{P}_{ab}(z,  \alpha_s) = D_{ab}(  \alpha_s)\delta(1-z) + K_{ab}( \alpha_s)\frac{1}{(1-z)_{+}} + R_{ab}(z, \alpha_s)  \; .
\label{Eq:Pdecomp}
\end{equation}

The coefficients $D$ and $K$ can be written as
$D_{ab}(\alpha_s) = \delta_{ab}d_a(\alpha_s)$, $K_{ab}(\alpha_s) = \delta_{ab}k_a(\alpha_s)$ and the coefficients $R_{ab}$ contain only  terms which are not singular for $z\rightarrow 1$. 
Each of those three coefficients can be expanded in powers of \as :
 \begin{equation}
 d_a(\alpha_s)=\sum_{n=1}^{\infty} \left( \frac{\alpha_s}{2\pi}\right)^n d_a^{(n-1)}, \;
  k_a(\alpha_s)=\sum_{n=1}^{\infty} \left( \frac{\alpha_s}{2\pi}\right)^n k_a^{(n-1)}, \;
   R_{ab}(z,\alpha_s)=\sum_{n=1}^{\infty} \left( \frac{\alpha_s}{2\pi}\right)^n R_{ab}^{(n-1)}(z)
\end{equation}
The plus-prescription and the $D$  part in Eq.~(\ref{Eq:Pdecomp}) can be expanded and Eq.~(\ref{EvolEq}) can be reformulated introducing a Sudakov form factor  $\Delta^S_a (  \mu^2  )$ 
(for details on the calculation see Appendix~\ref{appendix1})
which  is defined as:
\begin{equation}
\label{sud-def}
\Delta^S_a ( \mu^2 , \mu^2_0, \epsilon ) =  
\exp\left( -\int_{\mu_0^2}^{\mu^2}\frac
{d q^{2}}{q^{2}} 
\left[ \int_0^{z_M} k_a(\alpha_s) \frac{1}{1-z} {d}z  - d_a(\alpha_s)\right]\right)\; ,
\end{equation}
where an upper limit $z_M=1-\epsilon \;$ is introduced to allow numerical integration over $z$. To correctly reproduce the DGLAP evolution, $\epsilon \to 0$ is required. Note that the expression of $\Delta^S_a$ is different from the one used in Ref.~\cite{Hautmann:2017fcj} which is based on momentum weighted parton distributions; different forms of the Sudakov form factor are discussed in appendix~\ref{appendix1}.

The evolution equation for the parton density $ { f}_a(x,\mu^2)  $ at scale $\mu$ is then given by (as a solution of Eq.~(\ref{EvolEq}), see also \cite{Ellis:1991qj}):
\begin{equation}
\label{sudintegral2}
  { f}_a(x,\mu^2)  =  \Delta^S_a (  \mu^2  ) \  { f}_a(x,\mu^2_0)  
+ \sum_b
\int^{\mu^2}_{\mu^2_0} 
{{d q^{2} } 
\over q^{ 2} } 
{
{\Delta^S_a (  \mu^2  )} 
 \over 
{\Delta^S_a( q^2}  
 ) }
\int_x^{z_M} {\frac{dz}{z}} \;
\hat{P}_{ab} (\as , z) 
\;{f}_b\left({\frac{x}{z}},
q^{ 2}\right) 
\end{equation}
with the unregularized splitting functions $\hat{P}_{ab}$ (without the  $D_{ab}$ piece, replacing $1/(1-z)_+$ by $1/(1-z)$) and  $\mu_0$ being the starting scale.  The scale in \as\ can have different forms: it can be the evolution scale $q$  or it can be the transverse momentum $\qt=q(1-z)$ as described in Ref.~\cite{Martinez:2018jxt}.

In the \PBM -method, the evolution equation is solved iteratively, and each individual branching vertex is accessible, allowing the calculation of the transverse momenta (\qt) of the emitted partons. Further details on the formulation for TMD parton distributions are given in Ref.~\cite{Hautmann:2017fcj}).

At the collinear level, it was shown in Ref.~\cite{Hautmann:2017fcj} that the \PBM -approach reproduces the DGLAP evolution of parton densities exactly  \cite{Botje:2010ay}, if the renormalization scale (the argument in $\as$) is set to the evolution scale $q$ and if $z_M \to 1$. In Ref.~\cite{Martinez:2018jxt} the \PBM\ parton distributions are obtained from a fit~\cite{xFitterDevelopersTeam:2022koz,Alekhin:2014irh} of the parameters of the $x$-dependent starting distributions to describe high-precision deep-inelastic scattering data~\cite{Abramowicz:2015mha}. Two different parton distribution sets (we use \PBset\ as a shorthand notation for PB-NLO-HERAI+II-2018) were obtained,  \PBset~Set1, which for collinear distributions, agrees exactly with HERAPDF2.0NLO \cite{Abramowicz:2015mha}, and another set, \PBset~Set2, which uses \qt\ as the argument in \as, inspired by angular ordering conditions.  All \PBM\  parton distributions (and many others) are accessible in TMDlib and via the graphical interface TMDplotter~\cite{Abdulov:2021ivr,Hautmann:2014kza}.

In the following, we concentrate on the \PBset~Set1 scenario because of its direct correspondence to standard DGLAP solutions. For a discussion on \PBset Set~2 see Ref.~\cite{TaheriMonfared:2023ema,Bubanja:2023nrd}.

\section{Effect of \softgluon\  emissions in DGLAP equation (forward evolution)}
\label{sec:Sudakov}
The concept of resolvable and non-resolvable branchings with Sudakov form factors allows for an intuitive interpretation of the parton evolution pattern.  Sudakov form factors provide the probability to evolve from one scale to another scale without resolvable branching. 
While the concept of the  \PBM -method is similar to a parton shower approach, here, the method is used to solve the DGLAP evolution equation.

In order to illustrate the importance of resolvable and non-resolvable branchings we separate the Sudakov form factor $\Delta^S_a$
into a perturbative ($\qt > q_0$) and non-perturbative ($\qt < q_0$) part by introducing a resolution scale  $\zdyn = 1 - q_0/q$ (see Ref.~\cite{Hautmann:2019biw}).  This scale is motivated by angular ordering and the requirement to resolve an emitted parton with $\qt=q(1-z) > q_0$. 
Please note, in parton shower language the dynamical resolution scale $\zdyn$ is often identified with phase-space restrictions since the non-perturbaitve part is neglected.

The Sudakov form factor is then given by\footnote{It can be shown, that  $\Delta_a^{(\text{P})}$ coincides with the Sudakov form factor used in CSS~\cite{Collins:1984kg} up to next-to-leading and even partially next-to-next-to-leading logarithms (see \cite{vanKampen:2021oxe,PB-NNLL}). The non-perturbative Sudakov form factor $\Delta_a^{(\text{NP})}$ has a similar structure as the non-perturbative Sudakov form factor in CSS with the typical $\log (\mu^2/\mu_0^2)$ dependence.}
:
\begin{eqnarray}
\label{eq:divided_sud}
\Delta_a^S ( \mu^2 , \mu^2_0,\epsilon ) = && 
\exp \left(  -   
\int^{\mu^2}_{\mu^2_0} 
{{d q^{2} } 
\over q^{2} } \left[
 \int_0^{\zdyn(q)} dz 
  \frac{k_a(\as)}{1-z} 
- d_a(\as) \right]\right)\nonumber \\
 && \times  \exp \left(  -   
\int^{\mu^2}_{\mu^2_0} 
{{d q^{2} } 
\over q^{ 2} } 
 \int_{\zdyn(q)}^{z_M} dz 
  \frac{k_a(\as)}{1-z} 
\right) \nonumber \\
& = &  \Delta_a^{(\text{P})}\left(\mu^2,\mu_0^2,q^2_0 \right)  \cdot \Delta_a^{(\text{NP})}\left(\mu^2,\mu_0^2,\epsilon, q_0^2\right) \; ,
\end{eqnarray}
with $\epsilon$  defined via $z_M = 1 - \epsilon$.
Interestingly, $\Delta_a^{(\text{NP})}$ develops a $\log q$ dependence, and under certain conditions, $\Delta_a^{(\text{NP})}$ can even be calculated analytically (for example if $\as(q)$ is applied).

\begin{figure}[hbt!]
\centering
\includegraphics[width=0.45\textwidth]{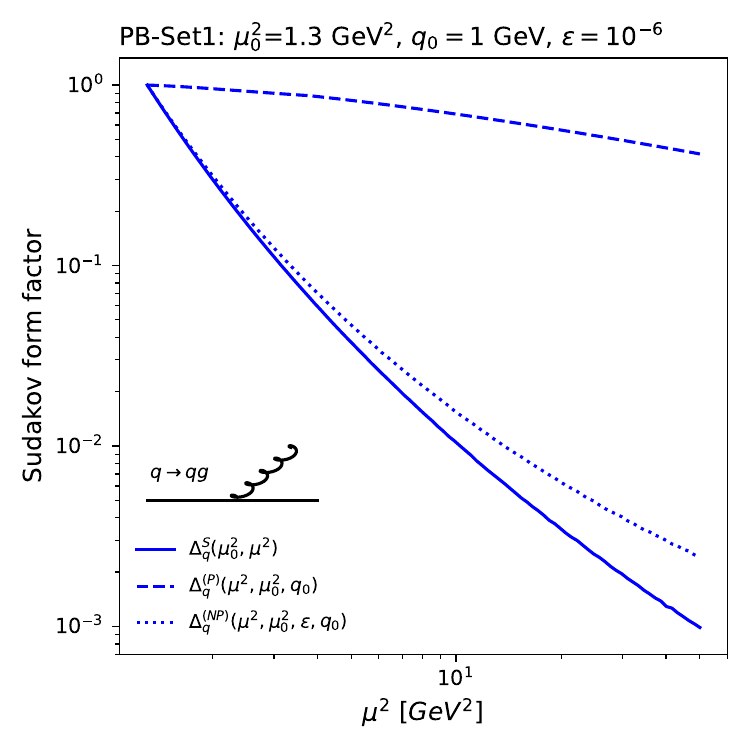}
\caption{Decomposition of the Sudakov form factor (solid line) into perturbative (dashed line) and non-perturbative (dotted line) emissions for a gluon emitted from a quark. Emissions with $q_{t}>q_{0}$ ($q_{t}<q_{0}$) are considered perturbative (non-perturbative).}\label{fig:pbset1_sudakov_decomposed}

\end{figure}

To study the contribution of non-perturbative and perturbative emissions in the Sudakov form factor, we use \PBset ~Set1  and calculate the Sudakov form factor for the $q\rightarrow qg$ case with a resolution parameter of $q_{0}=1$~\GeV\footnote{The value of $q_0$ is arbitrary, and chosen as $q_0 = 1 $ GeV for illustration only.}. As shown in Fig.~\ref{fig:pbset1_sudakov_decomposed}, the main contribution to the radiation probability ($1-\Delta_a^S ( \mu^2 , \mu^2_0 )$) is emissions with $q_{t}<q_{0}$. This highlights the significance of non-perturbative emissions in parton evolution.

In the following Subsections (\ref{subsec:pdf} and ~\ref{subsec:TMD}), we investigate the impact of non-resolvable splittings on collinear and TMD parton distribution functions. The effect on the DIS cross section calculation is discussed in Appendix~\ref{subsec:cross-section}.

\subsection{Collinear parton distribution functions}\label{subsec:pdf}

\begin{figure}[hbt!]
\begin{center} 
\includegraphics[width=0.32\textwidth]{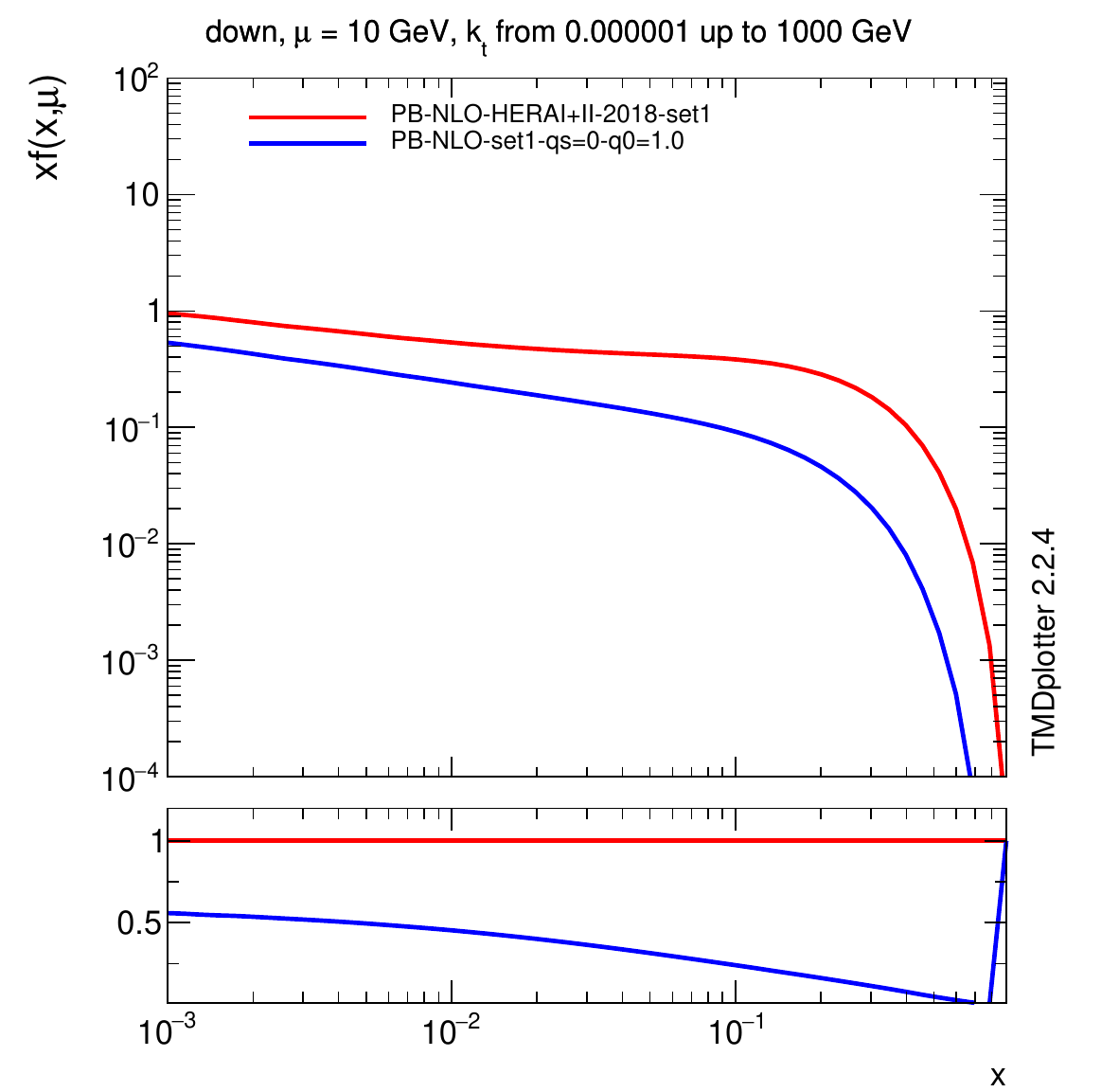}
\includegraphics[width=0.32\textwidth]{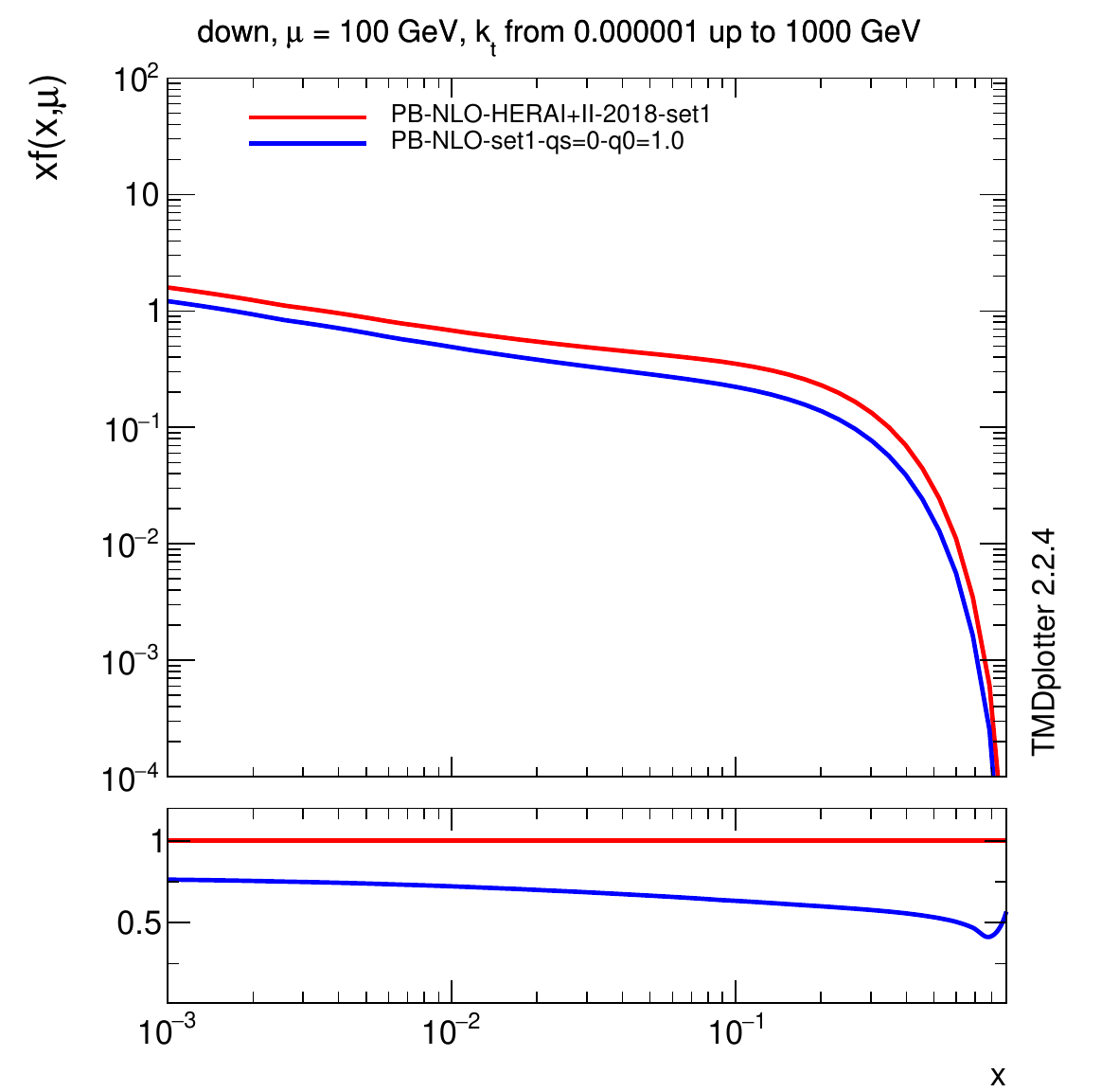}
\includegraphics[width=0.32\textwidth]{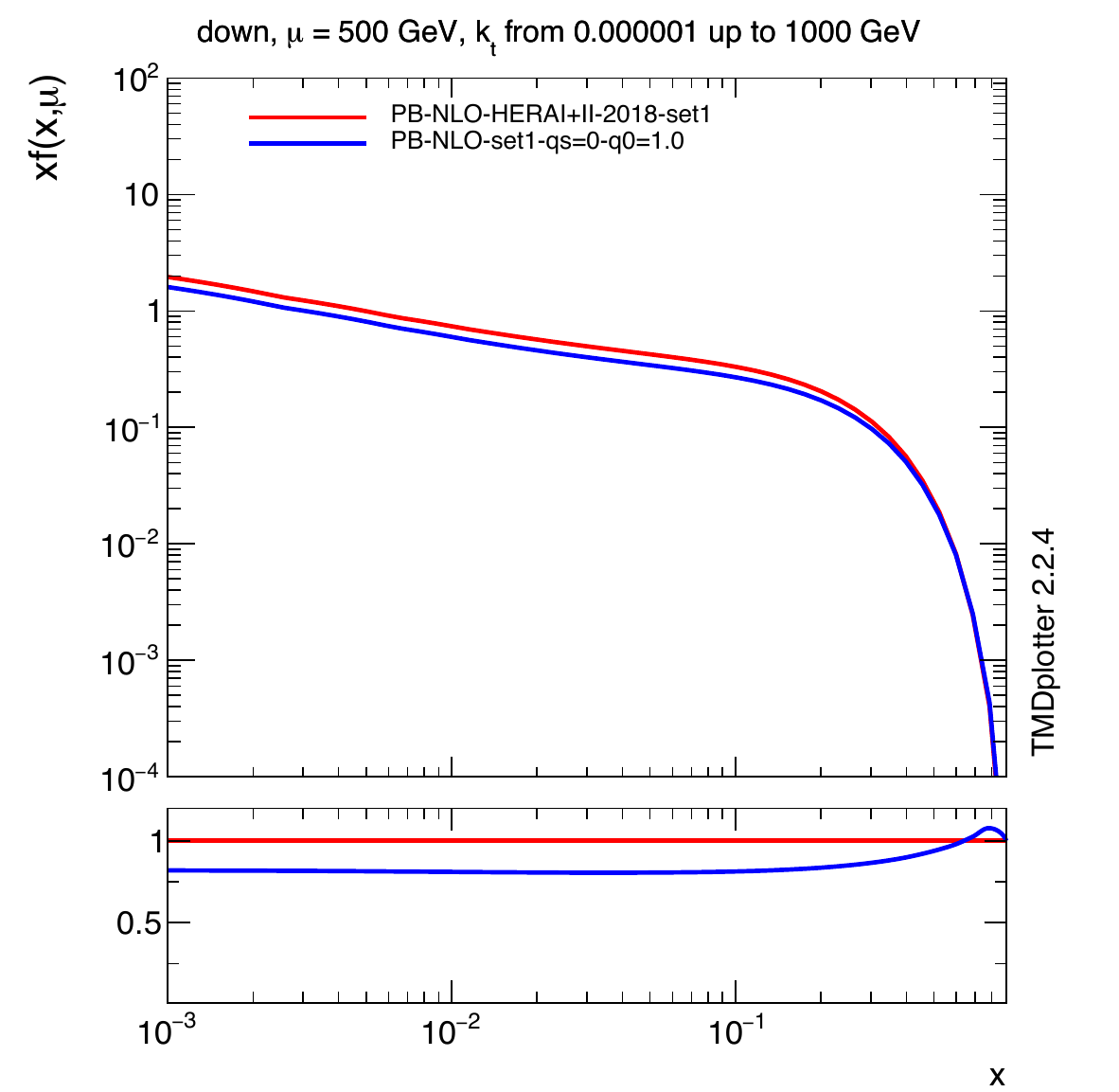}
\caption{\small Integrated down-quark distributions at $\mu  = 10, 100$~GeV and $\mu = 500$~GeV  obtained from the \protect\PBM -approach for different values of $z_M$: PB-NLO-HERAI+II-set1 applies $z_M \to 1$ and PB-NLO-set1 applies $z_M=\zdyn$ with $q_0=1$~GeV and without intrinsic -\kt\ distribution ($q_s=0$). The ratio plots show the ratios to the one for $z_M \to 1$. 
}
\label{PB-DGLAP}
\end{center}
\end{figure}

The influence of the non-perturbative Sudakov form factor on collinear PDFs is crucial, yet it has rarely been explicitly examined, as it is inherently included by construction in  solutions of the DGLAP evolution equation. In order to illustrate this, Fig.~\ref{PB-DGLAP} shows parton distributions obtained with the \PBM\ approach using the starting distributions from \PBset ~Set1 from Ref.~\cite{Martinez:2018jxt} for different scales $\mu$. 
We show distributions for down quark parton densities for different values of $z_M$: $z_M \to 1$ (default) and $z_M=\zdyn = 1 - q_0/q$, with $q_0 = 1 $ \GeV~\footnote{The starting parameters are the same as for \PBset~Set1, and the evoultion to higher $\mu$ is caluclated for different values of $z_M$.}. The results are shown without any intrinsic \kt\ distribution ($q_s=0$), which, by definition, does not affect the integrated parton densities.

The distributions obtained from \PBset~set1 with $z_M \to 1 $ are significantly different from those applying $z_M=\zdyn $, highlighting the importance of \softgluon\  contributions even for collinear distributions. By comparing the collinear distributions at low and high scales $\mu$, a clear scale dependence of the contribution from \softgluon s (and $\Delta_a^{(\text{NP})}$) becomes evident.

It is apparent that limiting the $z$-integration by $\zdyn$ (and neglecting $\Delta_a^{(\text{NP})}$) results in distributions that are no longer consistent with the collinear $\overline{\rm MS}$ factorization scheme. A similar conclusion was reached in Ref.~\cite{Nagy:2020gjv}.

\subsection{Transverse momentum distribution functions}\label{subsec:TMD}

\begin{figure}[hbt!]
\begin{center} 
\includegraphics[width=0.32\textwidth]{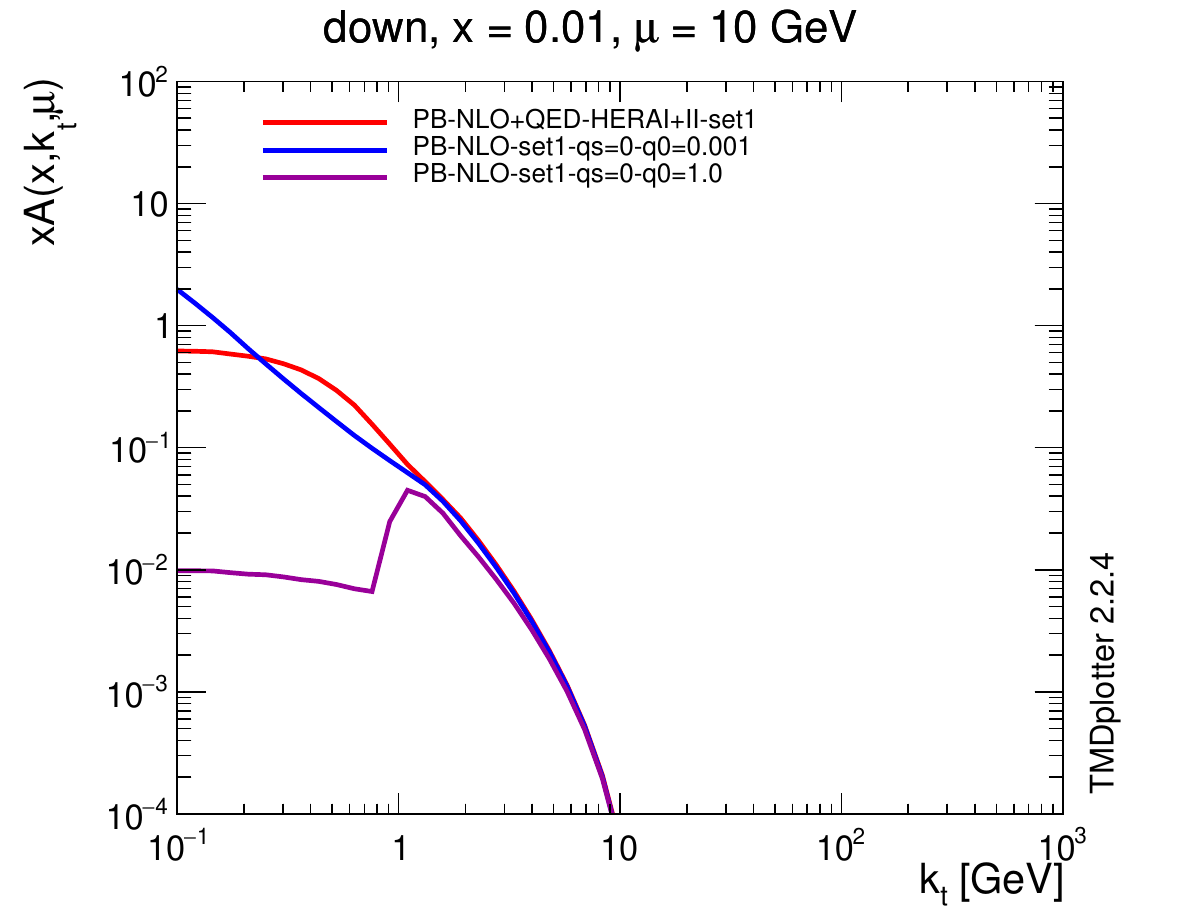}
\includegraphics[width=0.32\textwidth]{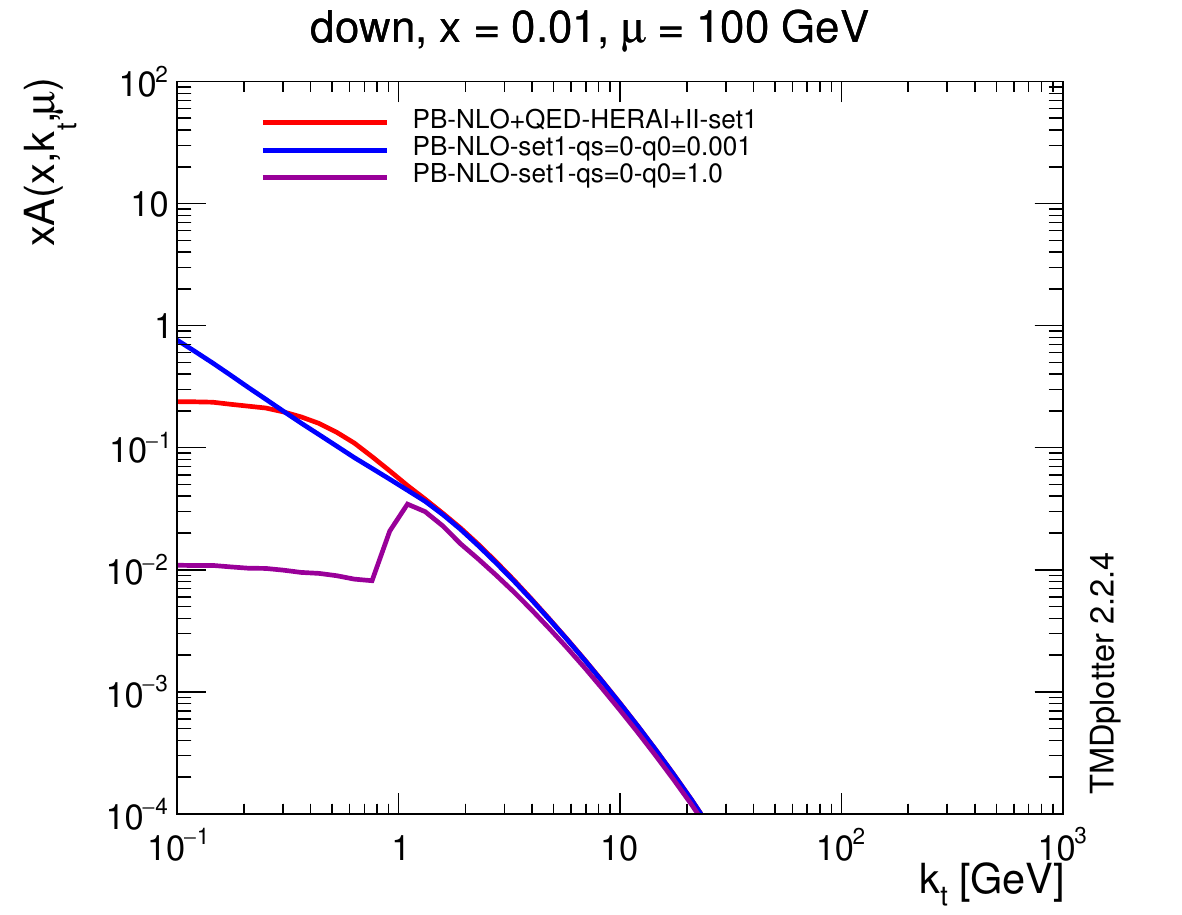}
\includegraphics[width=0.32\textwidth]{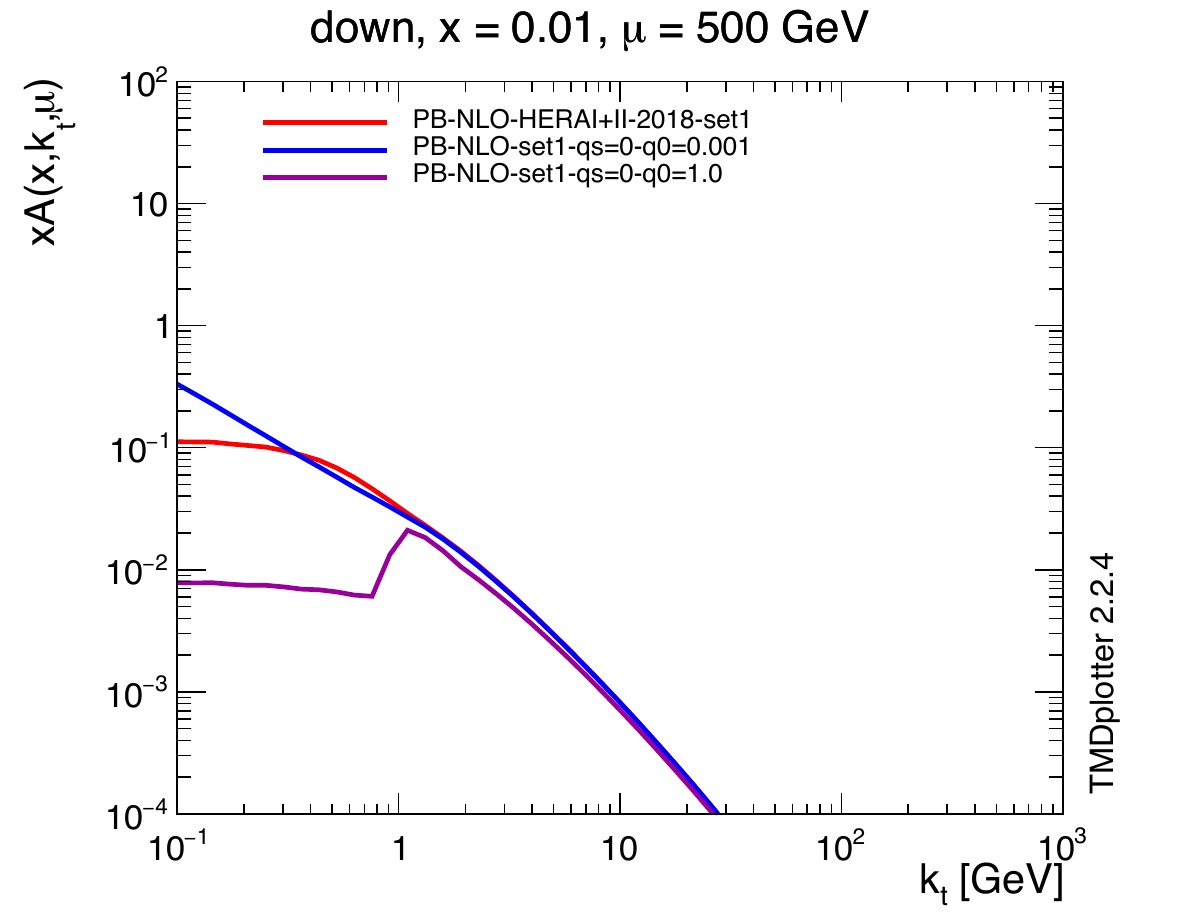}
\caption{\small Transverse momentum distributions of down quarks at $\mu  = 10, 100$~GeV (left, middle column) and $\mu = 500$ GeV (right column) obtained from the \protect\PBM -approach for $z_M \to 1$ as well as $z_M=\zdyn = 1 - q_0/q$. The red curve shows  \protect\PBset ~Set1 (including intrinsic-\kt ), the blue curve shows a prediction without including any intrinsic-\kt\ distribution ($q_s=0$), and the magenta curve shows a prediction applying $z_M=\zdyn$ with $q_0=1.0$ GeV without including intrinsic-\kt\ distributions.
 }
\label{TMD-DGLAP}
\end{center}
\end{figure}

Given that the \PBM -approach is also capable of determining TMD parton distributions (see Ref.~\cite{Hautmann:2017fcj}), we illustrate in Fig.\ref{TMD-DGLAP} the effect of the $z_M$ cut-off on the transverse momentum distribution. Specifically, we show results obtained with the \PBM -approach for down quarks using \PBset~Set1 with a default Gaussian width $q_s=0.5$~\GeV\ for the intrinsic \kt\ distribution. To focus solely on the evolution (as given in Eq.~(\ref{eq:divided_sud})), we also present results where no intrinsic \kt\ distribution is applied ($q_s=0$~\GeV; practically, a Gaussian distribution with $q_s=0.0001$~\GeV\ is used). Additionally, we demonstrate the impact of neglecting $\Delta_a^{(\text{NP})}$ by applying $\zdyn = 1 - q_0/q$, as in Fig.~\ref{PB-DGLAP}.

The transverse momentum distributions show very clearly the effect of $\Delta_a^{(\text{NP})}$. 

Applying the cut-off scale $z_M=\zdyn = 1 - q_0/q$ suppresses emissions with $\qt < q_0$ (though low-\kt\ contributions remain due to the vectorial addition of all intermediate emissions). However, very soft emissions are automatically included when $z_M \to 1$. Please note, that if $\alpha_s(\qt)$ is chosen a special treatment is needed for  $\qt < q_0$. As seen in Figs.~\ref{PB-DGLAP},\ref{TMD-DGLAP} the contribution of \softgluon s is $\mu$-scale dependent.

It is important to stress, that the non-perturbative Sudakov form factor is fixed from the fit to inclusive structure function measurement, which is not sensitive to transverse distributions. 
The non-perturbative Sudakov form factor significantly impacts the small-\kt\ region, where the intrinsic-\kt\ distribution also plays a crucial role. While the intrinsic-\kt\ distribution is expected to depend only on the hadron and not on the hard process, and hence be independent of the scale $\mu$, the non-perturbative Sudakov form factor varies with the scale of the hard process. The scale-dependent non-perturbative Sudakov form factor is derived from inclusive distributions, whereas the intrinsic-\kt\ distribution can only be constrained by processes sensitive to \kt, such as the \ptll-spectrum of Drell-Yan (DY) lepton pairs.
This non-trivial interplay between intrinsic-\kt\ distribution and non-perturbative Sudakov form factor allows for a consistent description of the DY \ptll -spectrum at small transverse momenta (discussed in Subsection~\ref{smallPtDY}).

\section{Effect of \softgluon\   emissions in initial state parton shower (backward evolution): from parton to hadron spectra }
\label{sec:Implications}
In Monte Carlo event generators, the initial parton shower is generated using a backward evolution approach for efficiency reasons (see e.g. Refs.~\cite{Bengtsson:1986gz,Sjostrand:1985xi,Marchesini:1987cf,Webber:1986mc,Marchesini:1983bm,Baranov:2021uol}). 
In event generators based on collinear parton densities, the accumulated transverse momentum of the initial state cascades determines the total transverse momentum of the hard process. For example, in DY production, initial state radiation determines the DY \pt . 
These generators simulate resolvable parton radiation in the initial state parton shower with collinear parton densities. Resolvable parton radiation is identified as radiation from a parton with transverse momentum exceeding a certain cutoff.
This is achieved either through the angular ordering condition and parameter $Q_g$ in \herwig\ ~\cite{Bahr:2008pv}[p~659] or through $z_{max}(Q^2)$ in \pythia\ ~\cite{Bierlich:2022pfr}[p~60]. These cuts on $z$ effectively remove $ \Delta_a^{(\text{NP})}$ from Eq.~(\ref{eq:divided_sud}).

In \cascade , a Monte Carlo generator based on TMD distributions, the transverse momentum of the hard process is already set by the TMD distribution. Consequently, the initial state parton shower is not allowed to change this momentum but can only add radiated partons. The initial state parton shower,  guided by the TMDs in a backward evolution, follows the same kinematic restrictions, as imposed in the parton densities, especially using $z_M \to 1$. Therefore, this approach allows us to study the impact of varying \zdyn\ values in the initial state shower without changing the overall kinematics. We use DY production at NLO (simulated by \MCatNLO ) supplemented with TMD distributions and initial state parton shower from \cascade , as described in Refs~\cite{Yang:2022qgk,Abdulhamid:2021xtt,Martinez:2020fzs,Martinez:2019mwt}. A comparison of parton showers used in \herwig\ and \cascade~\cite{Baranov:2021uol} is given in Ref.~\cite{Yang:2022qgk}.

In the following subsections, we study the impact of \softgluon\  emissions  on parton and hadron spectra using the backward evolution in the \cascade\ Monte Carlo event generator.

\subsection{\softgluon\  emissions in Drell-Yan \boldmath\pt -spectrum \label{smallPtDY}}
The DY \ptll -spectrum at large transverse momentum is described by hard single parton emissions. At lower \ptll, however, \softgluon s have to be resummed, and the intrinsic motion of partons inside the hadrons also plays a role. 
In Ref.~\cite{Martinez:2020fzs}, the DY \ptll -spectrum at low and high DY masses and at different center-of-mass energies is discussed. It is found that \PBset~Set2 provides a fairly reasonable description. 
In Ref.\cite{Bubanja:2023nrd}  a detailed analysis of the DY transverse momentum spectrum is presented  and it is found that after determining the parameter $q_s$ of the intrinsic-\kt\ distribution, \PBset~Set2 offers an excellent description of DY measurements across various DY masses and center-of-mass energies $\sqrt{s}$. 
The success of \PBset~Set2 in describing the DY \ptll -spectrum is attributed to the inclusion of $\Delta_a^{(\text{NP})}$ and the treatment of $\alpha_s$ at small scales.

\begin{figure}[hbt!]
\begin{center} 
\includegraphics[width=0.49\textwidth]{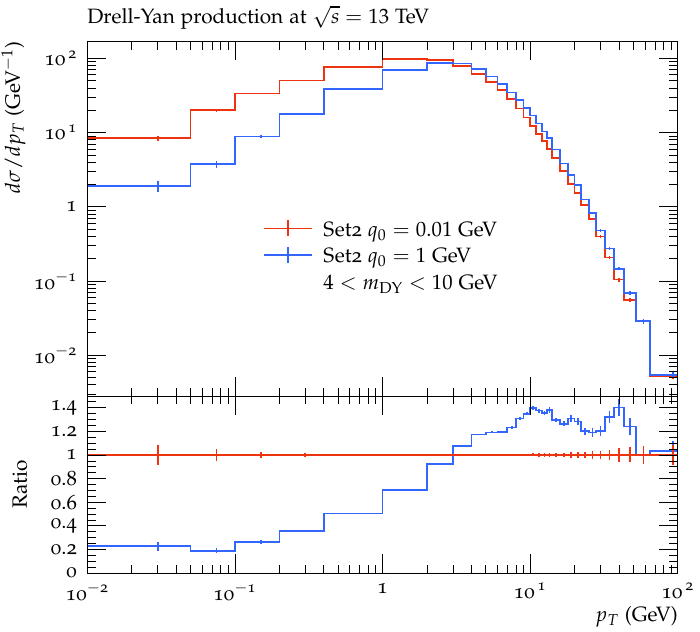}
\includegraphics[width=0.49\textwidth]{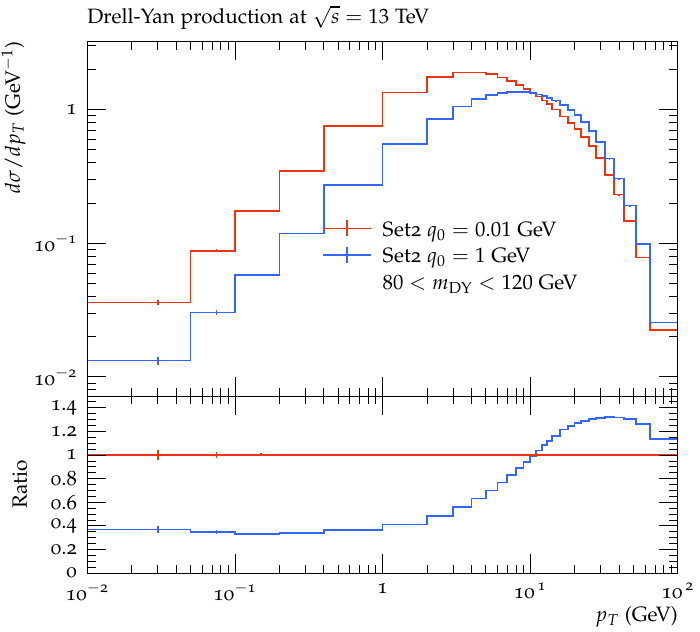}
\caption{\small The \ptll -spectrum of DY lepton pairs at low \mdy\ (left) and large \mdy\ (right) for different values of $q_0$ obtained with \mcatnlo\ matched to \cascade . }
\label{pt-DY}
\end{center}
\end{figure}

In Fig.~\ref{pt-DY} we show the \ptll -spectrum of DY lepton pairs at low \mdy\ and large \mdy\ at $\sqrt{s}=13$~\TeV\ obtained with \mcatnlo\ matched to \cascade\ for two scenarios: one with the default value of $q_0=0.01$~\GeV\ (as in  \PBset~Set2) and another one with $q_0=1$~\GeV. No intrinsic-\kt\ distribution is included. One can clearly see the  difference in the \pt -spectrum at low values between low \mdy\ and large \mdy : at low \mdy\ the ratio between the two scenarios is only 20\%, while at large \mdy\ the ratio is around 40\% . This illustrates the influence of the scale dependence of the non-perturbative Sudakov form factor (which is absent for $q_0=1$ \GeV ). This scale dependence implies that the intrinsic-\kt\ distribution must be also scale dependent, if $\Delta_a^{(\text{NP})}$ is neglected.

Standard parton shower Monte Carlo event generators based on collinear parton densities, which neglect the \softgluon\  contribution by requiring $Q_g$ or by imposing cuts on $z_{max}(Q^2)$, require an intrinsic-\kt\ spectrum that depends on $\sqrt{s}$. 
In Ref.~\cite{CMS-PAS-GEN-22-001}  a study on tuning the intrinsic-\kt\ distribution parameter for Monte Carlo event generators, \pythia\ and \herwig\ (with the most recent tunes) is reported. It is found that the Gaussian width $q_s$ of the intrinsic-\kt\ distribution varies with $\sqrt{s}$ for both generators and different tunes.

Applying the \PBM -method, it is shown in Ref.\cite{Bubanja:2024puv} that the energy dependence of the intrinsic-\kt\ distribution arises directly from neglecting \softgluon s (and the non-perturbative Sudakov form factor $\Delta_a^{(\text{NP})}$). This explanation addresses the longstanding question regarding the energy dependence of the intrinsic-\kt\ width and why a significantly large width (greater than expected from Fermi motion) was required at high $\sqrt{s}$~\cite{Huston:2004yp,Thome:2004sk}.

\subsection{\softgluon\  emissions in parton shower Monte Carlo event generators}
\begin{figure}[hbt!]
\begin{center} 
\includegraphics[width=0.49\textwidth]{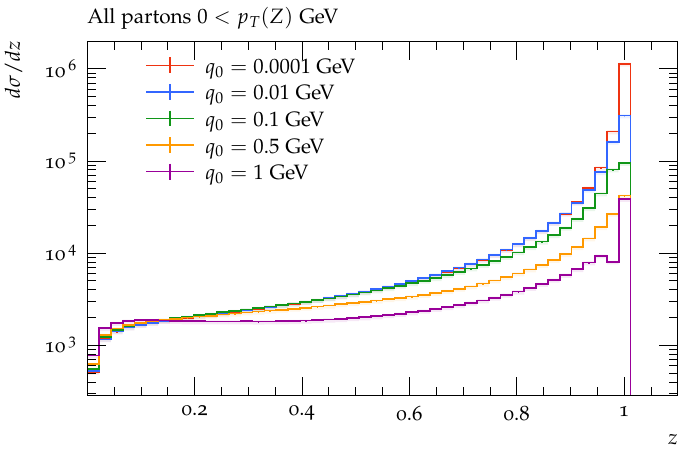}
\includegraphics[width=0.49\textwidth]{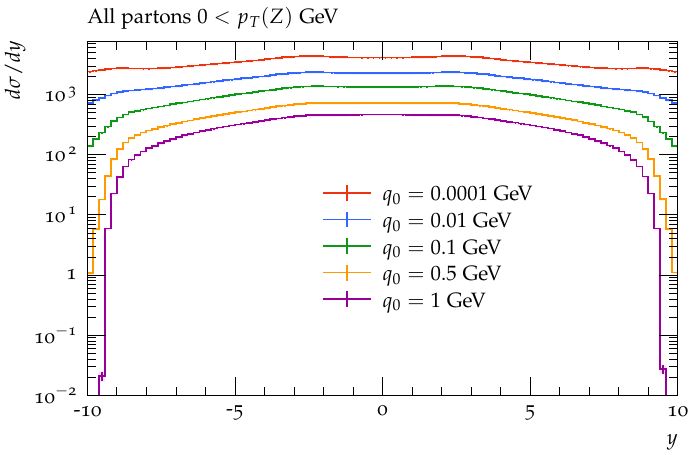}
\caption{\small Distributions of the splitting variable $z$ (left) and the rapidity of emitted partons $y$ (right) during the initial state shower for different values of $q_0$ in \PZ -boson events. }
\label{z-Zboson}
\end{center}
\end{figure}

We first investigate (Fig.~\ref{z-Zboson}) the spectrum of the splitting variable $z$ and the rapidity $y$ of emitted partons in the initial state parton shower for different values of $q_0$, which leads to different \zdyn\ values, with $\zdyn = 1 - q_0/q$ \footnote{We apply the cut \zdyn\ only in the parton shower.}. Since \zdyn\ depends on $q$ and very different values of $q=\qt/(1-z)$ are accessible during evolution, no clear cut in \zdyn\ is observed. However, the spectrum itself depends significantly on \zdyn\ and $q_0$. 

\begin{figure}[hbt!]
\begin{center} 
\includegraphics[width=0.49\textwidth]{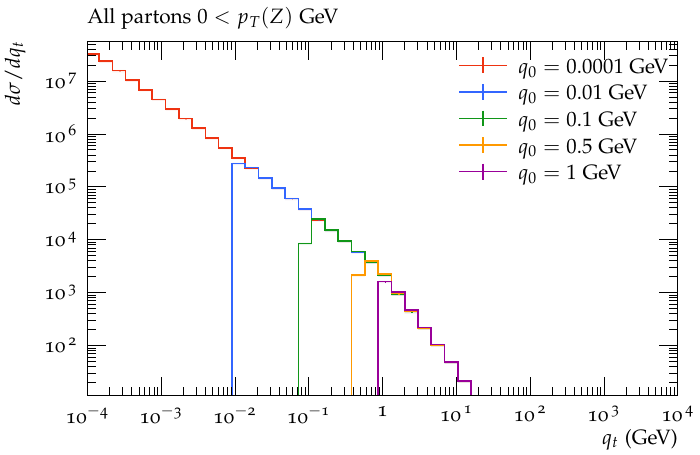}
\includegraphics[width=0.49\textwidth]{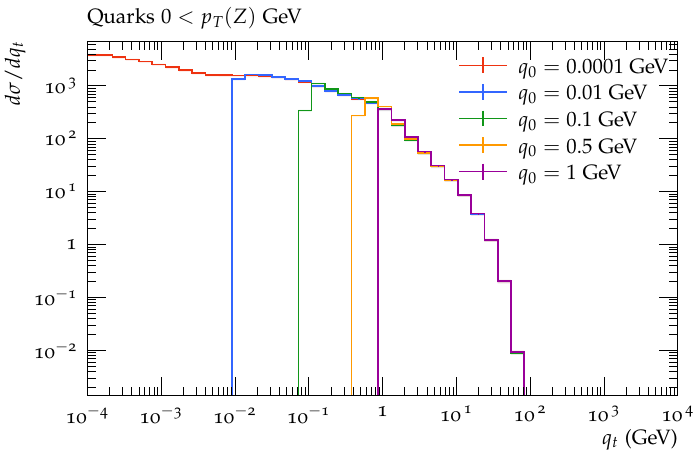}
\caption{\small Transverse momentum distributions of emitted partons  in the initial state cascade for different values of $q_0$ in  \PZ -boson events. On the left the spectrum for all partons is shown, on the right  the spectrum for quarks only is shown.
 }
\label{qt-Zboson}
\end{center}
\end{figure}

\begin{figure}[hbt!]
\begin{center} 
\includegraphics[width=0.49\textwidth]{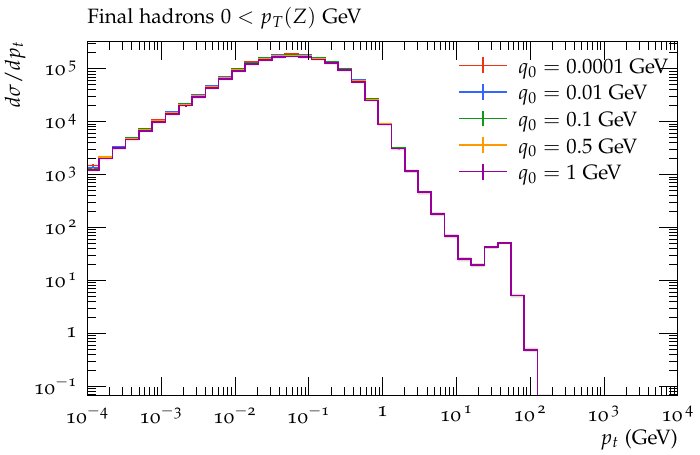}
\includegraphics[width=0.49\textwidth]{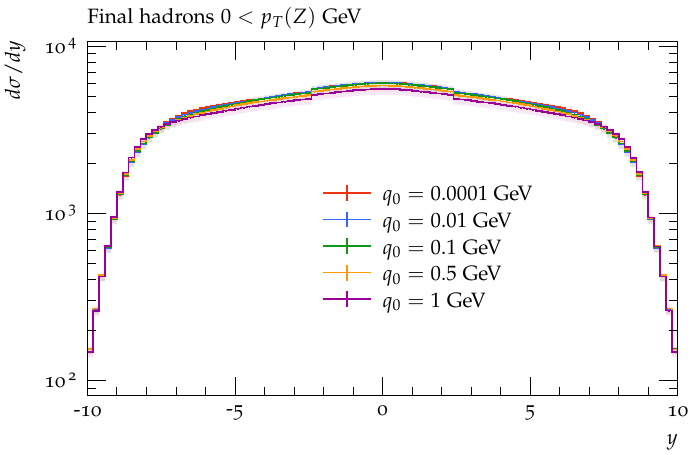}
\caption{\small Transverse momentum (left) and rapidity (right) distributions of particles for different values of $q_0$. The bump at large \pt\ comes from the muons of the \PZ -boson decay. }
\label{hadrons_qt-Zboson}
\end{center}
\end{figure}

Next, we analyze the transverse momentum spectrum of emitted partons in the initial state parton shower for varying $q_0$. In Fig.~\ref{qt-Zboson} (left) we show the transverse momentum (\qt ) spectrum of all partons emitted in the initial state shower for different values of $q_0$. 
Extremely low values of $q_0$ lead to a significant number of \softgluon\ emissions. In Fig.~\ref{qt-Zboson} (right), we display the same distributions but exclusively for emitted quarks. As anticipated, in such processes, where $g \to q\bar{q}$ and $q \to g q$, there is no singular behavior of the splitting function for $z \to 1$, and the spectrum at low transverse momenta is relatively flat compared to the case when gluon emission is included.
To determine whether these \softgluon s affect observable hadron distributions, we consider the Lund string hadronization model \cite{Bengtsson:1987kr,Bengtsson:1984yx,Sjostrand:1995iq,Sjostrand:2006za}, where gluons are treated as kinks in the color strings. Therefore, very \softgluon s are expected to have a negligible impact. In Fig.~\ref{hadrons_qt-Zboson}, we present the transverse momentum and rapidity spectra of particles in \PZ -boson events. The small dependence on $q_0$ observed in the rapidity spectrum is attributed to very low \pt -hadrons. While the \pt -spectrum of partons varies significantly with $q_0$ (as shown in Fig.\ref{qt-Zboson}(right)), there is essentially no observable effect on the final particle spectra, resulting in stable final results.
However, while \softgluon s can be accommodated in the Lund string model used in \cascade, their inclusion is not feasible in the \herwig\ cluster fragmentation model due to the requirement for a minimal fragmentation mass. The effect of \softgluon\ emissions, as treated by the NP-Sudakov form factor, cannot be mimicked by an additional Gauss distribution at the starting scale, since the number of emissions (perturbative and non-perturbative) depends on the available center-of-mass energy. This leads to a center-of-mass dependence of \softgluon\ emissions, although the NP-Sudakov form factor has no explicit center-of-mass dependence.

In summary, the effect of different \zdyn\ values on the non-perturbative Sudakov form factor $\Delta_a^{(\text{NP})}$ is quite significant for the low \pt -spectrum of partons in the initial state cascade (and thus important for DY \ptll ). However, this effect is negligible for final state hadron spectra and even less so for jets coming from the initial state shower.

\section{Summary and Conclusion}
\label{sec:concl} 
In this study, we have investigated the perturbative and non-perturbative regions of collinear parton densities by utilizing the DGLAP evolution equation reformulated in terms of Sudakov form factors, as applied in the \PBM -method. The separation of the Sudakov form factor into a perturbative and non-perturbative part is motived by investigations of the DY transverse momentum spectrum within the CSS approach.

Our findings reveal that soft and non-perturbative gluon emissions significantly impact inclusive parton distributions. These emissions are an essential part of the $\overline{\rm MS}$-scheme; neglecting those would lead to non-cancellation of important singular pieces.

With the requirement to describe and fit inclusive distributions, like inclusive DIS cross-sections, the non-perturbative Sudakov form factor is constrained and determined, once the factorization and evolution scheme for collinear parton densities is fixed (i.e. the choice of scale for the evolution of the strong coupling $\alpha_s$, which implies also a different form of the non-perturbative Sudakov form factor).

The non-perturbative Sudakov form factor, fixed from inclusive distributions, is essential for the description of the DY transverse momentum spectrum at small \ptll . The dependence on the number of emissions results in a width of the intrinsic-\kt\ distribution, which is process-independent, aligning with the expected Fermi motion of partons within hadrons. The proper treatment of \softgluon s and the non-perturbative Sudakov form factor leads to a solution of the longstanding issue with too large and scale-dependent width of the intrinsic-\kt\ distribution required by parton shower event generators.

We have also investigated the effect of \softgluon s on the transverse momentum spectrum of emitted partons and hadrons. Our results show that the treatment of \softgluon s is not important for observable hadrons in the final state. However, \softgluon s can be only included into hadronization models such as the Lund string model, while the \herwig\ cluster fragmentation requires a minimal mass, and thus requires the omission of \softgluon s.

In standard Mont Carlo event generators, \softgluon s can be included by lowering significantly the transverse momentum cut for emission during the parton shower.

By properly treating \softgluon s both in the evolution and in the parton shower, one can use standard parton densities and NLO coefficient functions without requiring new ones specific to the Monte Carlo event generator. The \PBM -method with its collinear and transverse momentum dependent distributions, and the TMD parton shower as implemented in \cascade, is currently the only approach that consistently treats \softgluon s and is applicable to NLO calculations.

\section{Appendix}

\subsection{Appendix: The Sudakov form factor \label{appendix1}}
The DGLAP splitting functions in the evolution equation Eq.~(\ref{EvolEq}) can be written in different forms, where the plus-prescription acts on different terms\footnote{In the derivation we only consider $\alpha_s(\mu)$}:
\begin{eqnarray}
{P}_{ab}(z,  \alpha_s) &=& {P}(z,  \alpha_s)_+ \label{Eq:PdecompPplus}\\
                                   &=&  D_{ab}(  \alpha_s)\delta(1-z) + K_{ab}( \alpha_s)\frac{1}{(1-z)_{+}} + R_{ab}(z, \alpha_s)~. \label{Eq:PdecompReal} \;
\end{eqnarray}
The plus prescription is given by:
\begin{eqnarray}
\int_0^1 dz \frac{\phi(z)}{(1-z)_+} &= & \int_0^1 dz \frac{\phi(z) - \phi(1)}{(1-z)} \\
 & = & \lim_{\epsilon \to 0}  \int_0^{1-\epsilon} dz \frac{\phi(z)}{(1-z)}  - \lim_{\epsilon \to 0}  \int_0^{1-\epsilon} dz \frac{\phi(1)}{(1-z)}  \;,
 \label{expandPlus}
\end{eqnarray}
where the last line is used for expanding the plus-prescription.

Starting from Eq.~(\ref{Eq:PdecompReal}) and Eq.~(\ref{expandPlus}) we obtain:
\begin{eqnarray}
\mu^2 \frac{{\partial }{ f}_a(x,\mu^2)}{{\partial } \mu^2}   & = &  \sum_b \int_x^{1} {\frac{dz}{z}} \; {P}_{ab} \left(z\right)  \;{f}_b\left({\frac{x}{z}}, \mu^{2}\right) \\
& = &  \sum_b \int_x^{1}  \frac{dz}{z} \; \left(D_{ab}\delta(1-z) + K_{ab}\frac{1}{(1-z)_{+}} + R_{ab}(z)  \right) \;f_b\left(\frac{x}{z} \right)\\
&=& \sum_b \int_x^{1}  \frac{dz}{z}  \left( \frac{K_{ab}}{1-z} + R_{ab}(z) \right)  f_b\left(\frac{x}{z} \right)  \nonumber \\
& & - f_a(x) \int_0^{1}  dz \left(\frac{k_{a}}{(1-z)} -d_{a}\delta(1-z) \right)~. \;
\end{eqnarray}

The Sudakov form factor $\Delta_S(\mu^2)$ is  defined as (see Eq.~(\ref{sud-def})):
\begin{equation}
\Delta_a^S(\mu^2,\mu^2_0,\epsilon) = \exp \left( - \int_{\mu_0^2}^{\mu^2} \frac{d q^{2}}{q^{2}} \int_0^{1-\epsilon} dz \left[   \frac{k_{a}}{(1-z)} -d_{a}\delta(1-z)  \right]
\right) \; . 
\label{Suda1}
\end{equation}

 It is important to note that the Sudakov form factor $\Delta^S_a(\mu^2)$ includes only the virtual parts of the diagonal splitting functions and, therefore, resums all the virtual corrections.
The evolution equation for $ {f_a\left(x,\mu^2\right)}/ {\Delta^S_a(\mu^2)}$ reads:
\newcommand{\Phat}{\ensuremath{\hat{P}}}
\begin{equation}
 \mu^2 \frac{\partial }{\partial  \mu^2} \frac{ f_a(x,\mu^2)}{\Delta^S_a(\mu^2)}   =  \sum_b \int_x^{1}  \frac{dz}{z} \; \Phat_{ab} \left(z\right) \;
 \frac{f_b\left({\frac{x}{z},\mu^2}\right)} {\Delta^S_a(\mu^2)}~.
\end{equation}

Instead of the above formulation one can also apply the plus prescription for the whole splitting function (\as\ in Eq.~(\ref{Eq:PdecompPplus})), keeping in mind that it applies only for the diagonal terms. With this we obtain:
\begin{eqnarray}
 \mu^2 \frac{{\partial }{ f}_a(x,\mu^2)}{{\partial } \mu^2}   & = &  \sum_b \int_x^{1} {\frac{dz}{z}} \; {P}_{ab} \left(z\right)  \;{f}_b\left({\frac{x}{z}}, \mu^{2}\right) \\
 & = &  \sum_b \int_0^{1}  \frac{dz}{z} \; P_{ab} \left(z\right) \;f_b\left({\frac{x}{z}}\right) 
  - \sum_b \int_0^{x}  \frac{dz}{z} \; P_{ab} (z)  \; f_b\left(\frac{x}{z}\right) \\
  & = &  \int_0^{1}  dz \; \Phat_{aa} \left(z\right)\left( \frac{1}{z}  \;f_a\left({\frac{x}{z}}\right) - f_a(z)\right) \\ \nonumber
 & &  +  \int_0^{1}  dz \; \Phat_{ab} \left(z\right) \frac{1}{z}  \;f_b\left({\frac{x}{z}}\right) 
 - \sum_b \int_0^{x}  \frac{dz}{z} \; \Phat_{ab} \left(z\right) \;f_b\left(\frac{x}{z}\right) \\
 & = &  \sum_b \int_x^{1}  \frac{dz}{z} \; \Phat_{ab} \left(z\right) \;f_b\left({\frac{x}{z}}\right)   - f_a(x)  \int_0^{1}  dz \; \Phat_{aa} \left(z\right)   ~,
\end{eqnarray}
where we dropped $\alphas$ and $\mu^2$ dependence for better readability. The unregularized splitting function is denoted by \Phat\  
(without the  $D_{ab}$ piece, and replacing $1/(1-z)_+$ by $1/(1-z)$), and the Sudakov form factor  $\Delta_a (\mu^2)$ is given by  (see \cite{Ellis:1991qj}):
\begin{equation}
\Delta_a(\mu^2)= \exp\left( -  \int_{\mu_0^2}^{\mu^2} \frac{d \mu^{\prime 2}}{\mu^{\prime 2}} \int_0^{1-\epsilon} dz \Phat _{aa} (\alphas,z) \right)~.
\label{Suda2}
\end{equation}
With this, the evolution equation for $ {f_a(x,\mu^2)}/ {\Delta_a(\mu^2)}$  is given by:
\begin{equation}
 \mu^2 \frac{\partial }{\partial  \mu^2} \frac{ f_a(x,\mu^2)}{\Delta_a(\mu^2)}   =  \sum_b \int_x^{1}  \frac{dz}{z} \; \Phat_{ab} \left(z\right) \;
 \frac{f_b\left({\frac{x}{z},\mu^2}\right)} {\Delta_a(\mu^2)}~.
\end{equation}

\begin{figure}[hbt!]
\begin{center} 
\includegraphics[width=0.49\textwidth]{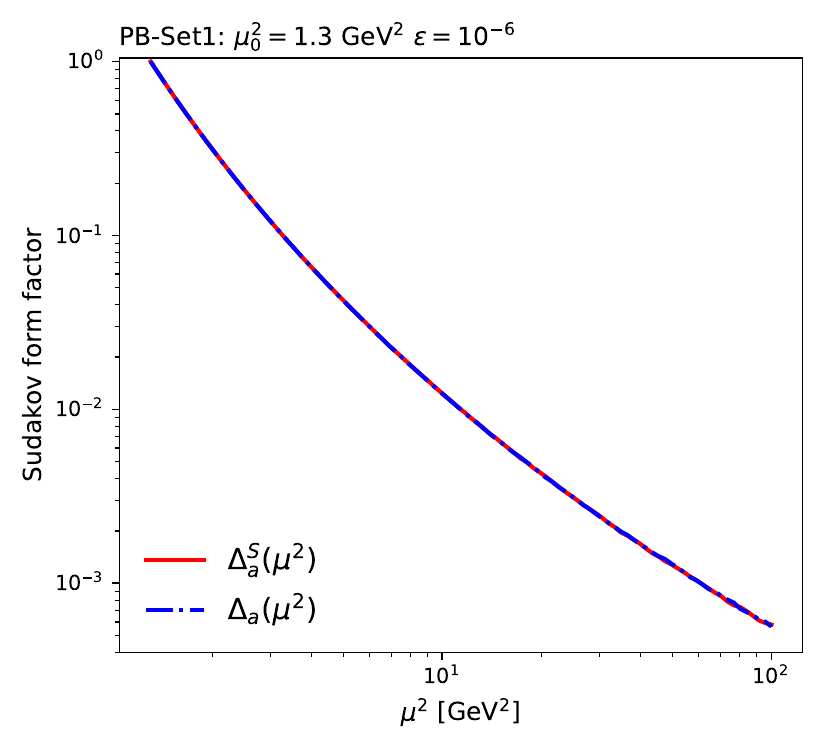}
\caption{\small Comparison of the Sudakov form factors obtained from Eq.~(\protect\ref{Suda1}) and Eq.~(\protect\ref{Suda2}). }
\label{SudaComp}
\end{center}
\end{figure}

The two Sudakov form factors $\Delta^S_a(\mu^2)$ and $\Delta_a(\mu^2)$ look different, but they are exactly the same, as also shown numerically in Fig.~\ref{SudaComp}.
Notably, only by using momentum weighted parton densities and applying the momentum sum rule for splitting functions can the Sudakov form factor be written in a form that includes all splitting functions (not just the diagonal ones), as shown explicitly in Ref.~\cite{Hautmann:2017fcj}.

\subsection{Appendix: Cross Section calculation}
\label{subsec:cross-section}
To ensure completeness, we include here a brief recap of the factorization of physical cross sections into scale-dependent parton densities and hard matrix elements (coefficient functions). For simplicity, we focus on the calculation of deep inelastic scattering using NLO parton densities and coefficient functions. The relevant expressions can be found in textbooks, such as in Ref.~\cite{Ellis:1991qj}, Eq.~4.80:
\begin{equation}
F_2(x,Q^2) = x \sum_{q,\bar{q}} e_q^2 \int_x^1 \frac{d\xi}{\xi} q(\xi,Q^2) \left[ \delta\left(1-\frac{x}{\xi} \right) + \frac{\alpha_s}{2 \pi} C_{\over{\rm MS}}\left( \frac{x}{\xi}\right) + \cdots  \right]~,
\label{F2NLO}
\end{equation}
with the coefficient function $C_{\over{\rm MS}}$ in the $\overline{\rm MS}$ scheme. The $\overline{\rm MS}$ coefficient function for massless quarks at ${\cal O}(\alpha_s)$ reads:
\begin{eqnarray}
C^{\over{\rm MS}}_q (z)& = & C_F \left[  2 \left(\frac{\log (1-z)}{1-z}\right)_+ - \frac{3}{2}\left( \frac{1}{1-z}\right)_+ - (1+z)\log(1-z) \right.  \nonumber \\
& &  \left. - \frac{1+z^2}{1-z} \log z + 3 + 2 z - \left( \frac{\pi^2}{3} + \frac{9}{2}\right) \delta(1-z) \right]~.
\end{eqnarray}

For a consistent formulation, the integral over $\xi$ in Eq.~(\ref{F2NLO}) must extend up to one, both in the expression for the cross section and in the expression for the parton density. Otherwise, singular pieces remain uncanceled. It becomes clear that the contribution of \softgluon\  emissions is important both in the parton densities and in the cross section calculations. Approaches where the integral is limited by $\zdyn $ lead to a different factorization scheme, in which the coefficient functions obtained in the collinear, massless $\overline{\rm MS}$ scheme are no longer appropriate. 
In Ref.~\cite{Frixione:2023ssx} \footnote{The preprint~\cite{Frixione:2023ssx} appeared a few days after this preprint.}, the same issue is discussed from the perspective of Monte Carlo event generators and the use of collinear parton densities in a backward evolution approach for the parton shower. By imposing a transverse momentum cut in the parton shower, a new factorization scheme needs to be developed, requiring new parton densities as well as new NLO coefficient functions, both of which will be Monte Carlo generator dependent.

From the above considerations, we can summarize, that the region of \softgluon\  emissions is extremely important both in the evolution of the parton densities and in the calculation of the hard cross section. 
Within the \PBM -approach, the non-perturbative Sudakov form factor is constrained by the fit to inclusive measurements to determine inclusive parton distributions: once the evolution frame is specified (depending on the scale choice for $\alpha_s$), the non-perturbative Sudakov form factor is fixed by the fit to inclusive measurements. The \PBM\ TMD distributions are then calculated without any further assumptions, in contrast to approaches like CSS, where the non-perturbative Sudakov form factor has to be determined separately by a fit to different distributions.

\bibliographystyle{mybibstyle-new.bst}
\raggedright  
\providecommand{\href}[2]{#2}\begingroup\raggedright\endgroup

\end{document}